# Emission of K - , L - and M - Auger Electrons from Cu Atoms

**Mohamed Assad Abdel-Raouf**

**Physics Department, Science College, UAEU,
Al Ain 17551, United Arab Emirates
assad@uaeu.ac.ae**

## Abstract

The emission of Auger electrons from the K-, L- and M-sheels of two configurations of copper atoms is investigated in detail within the frame work of the momentum average technique. Ab initio calculations for the bound state wavefunctions of the ionized atoms are performed using a Gaussian computer code. The continuum wavefunctions of the emitted electrons are determined by employing an effective potential variational approach. Slight differences have been noticed between the results of both configurations. The transition rates, energies and widths of all possible Auger electrons are listed in Tables. Slight differences have been noticed between the results of the two types of copper atoms.

-----------------------------------------------------------------------------------

# 1. Introduction

Inner shell ionizations of atoms and molecules could be initiated via electron, positron or photon interactions. In all cases holes are created followed by rearrangement processes. The most interesting process is the one in which an electron from an outer shell falls to fill a hole. The energy difference between the two states is absorbed by an electron belonging to a higher orbit. Immediately after absorption the electron leaves the interaction region yielding a doubly ionized atom. The transition rate, $A^{(0)}_a$, of the flying electron, usually referred to (after the discoverer of the phenomenon by Auger 1923 and 1925) as Auger electron, leads to valuable information about the electronic structure of the original system. This provides us with a plausible explanation for the vigorous interest shown by industrial institutions in the development and employment of Auger scanners. (For recent review of the data, see Safronova et al 2001). The main goal of the present work is to determine the Auger rates, energies and widths of all possible transitions from the K-, L- and M-shells of two configurations of copper atoms. For this purpose, the momentum average technique is employed, whilst the wavefunctions of the bound electrons are obtained using an ab initio Gaussian computer code. The continuum wavefunctions of the emitted electrons are evaluated by employing an effective potential variational approach suggested by Abdel-Raouf et al (1999, 2002) for solving the corresponding Schrödinger equation. In section 2 the mathematical formalism is shortly presented. Section 3 contains the results and discussions of our data. Concluding remarks are also given in the same section

## 2. Mathematical Formalism

Theoretically, the Auger transition rate $A^{(0)}_a$ is determined by (Wentzel 1927):

$$A^{(0)}_a = \frac{2\pi}{\tau} \left| <\varphi_3(r_1)\, \varphi_c(r_2) \mid \mid \frac{1}{r_1 - r_2} \mid \mid \varphi_1(r_1)\, \varphi_2(r_2) > \right|^2 , \qquad (1)$$

where $\tau = h\, a_0/2\pi e^2 = h^2/4\,\pi^2 me^4 = 2.42\times10^{-17}$ sec, is the time atomic unit, $\varphi_c$ is the wavefunction of the emitted Auger electron, (also referred to as the continuum wavefunction). $\varphi_3$ is the wavefunction describing the electron that filled the hole. The wavefunctions $\varphi_1$ and $\varphi_2$ represent the original bound states of these two electrons. They are determined in the present work using a Gaussian computer code. The wavefunction of the Auger electron $|\phi_C>$ is calculated by solving Schrödinger's equation

$$(H\text{-}E)|\phi_C> = |0> , \qquad (2)$$

where H is given by

$$H = \frac{-\hbar^2}{2m}\nabla_r^2 + V(r) . \qquad (3)$$

$V(r)$ is the effective potential seen by the continuum electron at a distance $r$ from the infinitely heavy nucleus of the doubly ionized atom.

Following McGuire (1972), in the angular momentum average scheme, the Auger rate for the transition $n_1\ell_1 n_2\ell_2 \rightarrow n_3\ell_3 n_c\ell_c$ is connected with the two electrons Auger rate $A^{(0)}_a$ by two different forms:

*Case I: $n_1\,\ell_1 \neq n_2\,\ell_2$*

$$\overline{A_a} = [(h_3+1)\,(4\,\ell_1+2\text{-}h_1)\,(4\ell_2+2-h_2)] \,/\, [(4\ell_1+2)\,(4\ell_2+2)]\; A^{(0)}_a(i \rightarrow f) , \qquad (4)$$

where $h_3$, $h_1$, and $h_2$ are the numbers of holes in the subshells $n_3 \ell_3$, $n_1 \ell_1$ and $n_2 \ell_2$, respectively. Also, the indices $i$ and $f$ specify an initial and a final state, respectively.

***Case II:***      $n_1 \ell_1 = n_2 \ell_2$

$$\overline{A_a} = [(h_3+1)(4\ell_1+2-h_1)(4\ell_1+2-h_1)] / [(4\ell_1+2)(4\ell_1+2)]\ A_a^{(0)}(i \to f),\quad (5)$$

**Where**

$$A_a^{(0)}(n_1\ell_1 n_2\ell_2 \to n_3\ell_3 n_c\ell_c) = (1/2)(2\ell_c+1)(2\ell_1+1)(2\ell_1+1)\sum_{h,g} hg[\mathrm{Im}(h,g)]^2$$

$$(6)$$

$h$ and $g$ are the spin and orbital quantum numbers . The corresponding Auger widths are determined by

$$\Gamma_a(i) = \sum_f \overline{A}_a(i \to f)$$

## 3. Results and Discussions

In this section we present the final results obtained after lengthy investigations of the problem under consideration. These are the Auger rates $\overline{A}_a$, Auger widths $\overline{\Gamma}_a$, and Auger energies $E_a$ characteristic to K - , L - and M - inner shell ionizations for two configurations of the copper atom, (to be referred to in the Tables as Cu I and Cu II).

The following three subsections contain the results and discussions of our Auger transitions calculated for the inner shell ionizations mentioned above for Cu I and Cu II.

### 3.1. K-Shell Auger Transitions

In Table 1 we list the Auger transition rates $\bar{A}_a$'s (measured in sec-1) corresponding the initial state $1s^1 2s^2 2p^6 3s^2 3p^6 3d^{10} 4s^1$ of Cu I. The Auger transitions corresponding to the initial state $1s^1 2s^2 2p^6 3s^2 3p^6 3d^9 4s^2$ of Cu II are given in Table 2. (Remember that the superscripts refer to the numbers of electrons in the occupied orbits). From the Tables we conclude the following remarks:

(1) The transitions $\bar{A}_a(1s \rightarrow 2s2s)$, $\bar{A}_a(1s \rightarrow 2s2p)$, $\bar{A}_a(1s \rightarrow 2p2p)$, and $\bar{A}_a(1s \rightarrow 2p3p)$ represent the dominant contributions to the Auger widths of the initial state. The third transition is the largest one.

(2) For Cu I and Cu II, the Auger rates in which the outer electron is involved (e.g. $\bar{A}_a(1s \rightarrow 3d4s)$ and $\bar{A}_a(1s \rightarrow 3d3d)$) provide small contributions to the Auger widths, whilst $\bar{A}_a(1s \rightarrow 3d4s)$ has the smallest contribution in case of Cu I and $\bar{A}_a(1s \rightarrow 4s4s)$ yields the smallest contribution for Cu II. The ratio of the Auger widths of Cu I and Cu II is 1.04878.

(3) The resultant transition energies for Cu I show that $E_c$ increases with n and ℓ for identical or different effective electrons. For example $E_c = 441.3$ Ry for the transition $\bar{A}_a(1s \rightarrow 2s2s)$, $E_c = 468.08$ Ry for the transition $\bar{A}_a(1s \rightarrow 2p2p)$ and $E_c = 593.38$ Ry for the transition $\bar{A}_a(1s \rightarrow 3s3s)$. $E_c = 600.67$ Ry for the transition $\bar{A}_a(1s \rightarrow 3p3p)$ and $E_c = 656.91$ Ry for the transition $\bar{A}_a(1s \rightarrow 3d3d)$. Similar behavior is concluded from Table 2.

**Table 1: Auger transition rates in sec$^{-1}$ for initial states of Cu (I) , the energy of the emitted electron (in Ry) and the angular momentum of this electron (c). The number between parentheses is the power of 10.**

| Cu (I) Initial State : $1s^12s^22p^63s^23p^63d^{10}4s^1$ | | | |
|---|---|---|---|
| **Final State** | **$\ell_c$** | **$E_c$ (Ry)** | **$\overline{A_a}$ (Sec$^{-1}$)** |
| $1s^22p^63s^23p^63d^{10}4s^1$ | 0 | 441.3 | 4.3299(13) |
| $1s^22s^22p^43s^23p^63d^{10}4s^1$ | 0 | 468.08 | 2.33219(13) |
| $1s^22s^22p^63s^23p^63d^{10}4s^1$ | 0 | 593.38 | 1.2530(12) |
| $1s^22s^22p^63s^23p^63d^{10}4s^1$ | 0 | 600.67 | 8.0733(11) |
| $1s^12s^22p^63s^23p^43d^84s^1$ | 0 | 656.91 | 5.2068(7) |
| | 2 | | 2.0806(8) |
| | 4 | | 1.3080(10) |
| $1s^22s^12p^63s^23p^63d^{10}4s^1$ | 1 | 451.66 | 1.7691(13) |
| $1s^22s^12p^63s^13p^63d^{10}4s^1$ | 0 | 516.39 | 8.1176(12) |
| $1s^22s^12p^63s^23p^53d^{10}4s^1$ | 1 | 520.97 | 1.821(13) |
| $1s^22s^12p^63s^23p^63d^94s^1$ | 2 | 571.97 | 1.1006(12) |
| $1s^22s^12p^63s^23p^63d^9$ | 0 | 548.35 | 2.1935(11) |
| $1s^22s^22p^53s^13p^63d^{10}4s^1$ | 1 | 526.72 | 1.5861(13) |
| $1s^22s^22p^53s^23p^53d^{10}4s^1$ | 0 | 530.4 | 1.3412(13) |
| $1s^22s^22p^53s^23p^63d^94s^1$ | 1 | 558.51 | 1.767(11) |
| | 3 | | 9.0737(12) |
| $1s^22s^22p^63s^23p^63d^{10}$ | 1 | 558.69 | 1.9985(11) |
| $1s^22s^22p^63s^13p^63d^{10}4s^1$ | 1 | 597.01 | 3.66659(12) |
| $1s^22s^22p^63s^13p^63d^94s^1$ | 2 | 625.15 | 1.170(11) |
| $1s^22s^22p^63s^13p^63d^{10}$ | 0 | 624.45 | 4.6045(10) |
| $1s^22s^22p^63s^23p^53d^94s^1$ | 1 | 628.79 | |
| $1s^22s^22p^63s^23p^53d^{10}$ | 1 | 628.10 | 5.1105(10) |
| $1s^22s^22p^63s^23p^63d^9$ | 2 | 656.2 | 1.3289(9) |
| | | | $\Gamma_a$=7.445(14) |

**Table 2: Auger transition rates in sec$^{-1}$ for CU (II), the energy of the emitted electron (in Ry) and the angular momentum of this electron (c). The number between parentheses is the power of 10.**

| Final State | $\ell_c$ | $E_c$ (Ry) | $\overline{A_a}$ (Sec$^{-1}$) |
|---|---|---|---|
| **Cu (II)** Initial State : $1s^12s^22p^63s^23p^63d^94s^2$ | | | |
| $1s^22p^63s^23p^63d^94s^2$ | **0** | **441.18** | 9.6605(13) |
| $1s^22s^22p^43s^23p^63d^94s^2$ | **0** | **461.96** | 2.33587(13) |
| | **2** | | 6.0259(12) |
| $1s^22s^22p^63p^63d^94s^2$ | **0** | **593.17** | 2.6648(12) |
| $1s^22s^22p^63s^23p^43d^94s^2$ | **0** | **600.44** | 5.9835(11) |
| | **2** | | 6.0259(12) |
| $1s^22s^22p^63s^23p^63d^74s^2$ | **0** | **655.5** | 2.8959(7) |
| | **2** | | 8.7430(7) |
| | **4** | | 1.1500(10) |
| $1s^22s^22p^63s^23p^63d^94s^2$ | **1** | **505.34** | 9.4169(13) |
| $1s^22s^12p^63s^13p^63d^94s^2$ | **0** | **565.32** | 5.5562(12) |
| $1s^22s^12p^63s^23p^53d^94s^2$ | **1** | **520.80** | 1.9904(13) |
| $1s^22s^22p^63s^23p^63d^84s^2$ | **2** | **548.35** | 6.0191(12) |
| $1s^22s^12p^63s^23p^63d^84s^2$ | **0** | **548.92** | 1.3819(11) |
| $1s^22s^22p^53s^13p^63d^94s^2$ | **1** | **527.50** | 1.5831(13) |
| $1s^22s^22p^53s^23p^53d^94s^2$ | **0** | **531.18** | 1.3312(13) |
| | **2** | | 1.4113(14) |
| $1s^22s^22p^53s^23p^63d^84s^2$ | **1** | **558.69** | 1.6522(11) |
| | **3** | | 8.1708(12) |
| $1s^22s^22p^53s^23p^63d^94s^1$ | **1** | **559.79** | 2.9211(11) |
| $1s^22s^22p^63s^13p^53d^94s^2$ | **1** | **596.79** | 2.747(12) |
| $1s^22s^22p^63s^23p^63d^94s^2$ | **2** | **624.33** | 7.7606(10) |
| $1s^22s^22p^63s^13p^63d^84s^1$ | **0** | **624.97** | 2.3182(10) |
| $1s^22s^22p^63s^23p^53d^84s^2$ | **1** | **627.97** | 1.8731(10) |
| | **3** | | 9.6887(11) |
| $1s^22s^22p^63s^13p^53d^84s^1$ | **1** | **628.65** | 5.0597(10) |
| $1s^22s^22p^63s^23p^63d^84s^1$ | **2** | **656.2** | 1.337(9) |
| $1s^22s^22p^63s^23p^63d^9$ | **0** | **656.8** | 7.828(8) |
| | | | $\Gamma_a$=7.0988(14) |

## 3.2. L -Shell Auger Transitions

In this case two holes are possible, namely a 2s – hole or a 2p – hole, leading to two sets of Auger transitions, (corresponding to two initial ionized states) in each configuration of the copper atom. Tables (3), (4) contain all possible transitions for Cu I ions and Tables (5), (6) demonstrate all possible Auger transitions in case of Cu II ions.

**Table 3: Auger transition rates in sec$^{-1}$ resulting from a 2s-hole in Cu (I) , the energy of the emitted electron (in Ry) and the angular momentum of this electron (c). The number between parentheses is the power of 10.**

| Cu (I) Initial State : $1s^2 2s^1 2p^6 3s^2 3p^6 3d^{10} 4s^1$ | | | |
|---|---|---|---|
| **Final State** | $\ell_c$ | $E_c$ (Ry) | $\overline{A_a}$ (Sec$^{-1}$) |
| $1s^2 2s^2 2p^5 3s^1 3p^6 3d^{10} 4s^1$ | 1 | 4.481 | 2.352 (11) |
| $1s^2 2s^2 2p^5 3s^2 3p^5 3d^{10} 4s^1$ | 0 | 8.93 | 4.245 (12) |
|  | 2 |  | 2.98572(12) |
| $1s^2 2s^2 2p^5 3s^2 3p^5 3d^9 4s^1$ | 1 | 11.23 | 8.321 (11) |
|  | 3 |  | 3.242 (12) |
| $1s^2 2s^2 2p^5 3s^2 3p^5 3d^{10}$ | 1 | 10.10 | 4.233 (10) |
| $1s^2 2s^2 2p^6 3s^0 3d^{10} 4s^1$ | 0 | 16.787 | 1.0942 (14) |
| $1s^2 2s^2 2p^6 3s^1 3p^5 3d^{10} 4s^1$ | 1 | 20.56 | 3.8243 (14) |
| $1s^2 2s^2 2p^6 3s^1 3p^6 3d^9 4s^1$ | 2 | 48.438 | 1.4607( 14) |
| $1s^2 2s^2 2p^6 3s^1 3p^6 3d^{10}$ | 0 | 48.622 | 2.1180 (13) |
| $1s^2 2s^2 2p^6 3s^2 3p^4 3d^{10} 4s^1$ | 0 | 24.767 | 2.6320(12) |
|  | 2 |  | 9.7623(12) |
| $1s^2 2s^2 2p^6 3s^2 3p^5 3d^9 4s^1$ | 1 | 52.156 | 1.9318 (13) |
|  | 3 |  | 1.9258 (14) |
| $1s^2 2s^2 2p^6 3s^2 3p^5 3d^{10}$ | 1 | 52.69 | 6.2562 (12) |
| $1s^2 2s^2 2p^6 3s^2 3p^6 3d^8 4s^1$ | 0 | 80.125 | 1.6110 (12) |
|  | 2 |  | 1.0980 (12) |
|  | 4 |  | 7.4922 (13) |
| $1s^2 2s^2 2p^6 3s^2 3p^6 3d^9$ | 2 | 80.289 | 1.2354(12) |
|  |  |  | $\Gamma_a$=7.6049 (14) |

**Table 4: Auger transition rates in sec⁻¹ resulting from a 2p-hole in Cu (I) , the energy of the emitted electron (in Ry) and the angular momentum of this electron (c). The number between parentheses is the power of 10.**

| Cu (I) Initial State : $1s^2 2s^2 2p^5 3s^2 3p^6 3d^{10} 4s^1$ | | | |
|---|---|---|---|
| **Final State** | $\ell_c$ | **$E_c$ (Ry)** | $\overline{A_a}$ **(Sec⁻¹)** |
| $1s^2 2s^2 2p^6 3p^6 3d^{10} 4s^1$ | 1 | 6.2862 | 9.1591(11) |
| $1s^2 2s^2 2p^6 3s^1 3p^5 3d^{10} 4s^1$ | 0 | 10.08 | 9.612(13) |
|  | 2 |  | 2.7327 (13) |
| $1s^2 2s^2 2p^6 3s^1 3p^6 3d^9 4s^1$ | 1 | 37.994 | 2.1257(13) |
|  | 3 |  | 1.1047(13) |
| $1s^2 2s^2 2p^6 3s^1 3p^6 3d^{10}$ | 1 | 38.188 | 2.7152(10) |
| $1s^2 2s^2 2p^6 3s23p^4 3d^{10} 4s^1$ | 1 | 14.313 | 3.3295 (14) |
|  | 3 |  | 2.0449(13) |
| $1s^2 2s^2 2p^6 3s^2 3p^5 3d^9 4s^1$ | 0 | 41.696 | 4.7235(13) |
|  | 2 |  | 2.7890(14) |
|  | 4 |  | 2.8723(13) |
| $1s^2 2s^2 2p^6 3s^2 3p^5 3d^{10}$ | 0 | 41.862 | 1.2230(12) |
|  | 2 |  | 7.4192(9) |
| $1s^2 2s^2 2p^6 3s^2 3p^6 3d^8 4s^1$ | 1 | 69.698 | 8.0434(12) |
|  | 3 |  | 1.4483(14) |
|  | 5 |  | 4.7290(12) |
| $1s^2 2s^2 2p^6 3s^2 3p^6 3d^9$ | 1 | 69.88 | 2.8787(10) |
|  | 3 |  | 5.5947(10) |
|  |  |  | $\Gamma_a$=1.02387 (15) |

**Table 5: Auger transition rates in sec⁻¹ resulting from a 2s-hole in Cu (II) , the energy of the emitted electron (in Ry) and the angular momentum of this electron (c). The number between parentheses is the power of 10.**

| Cu (II) Initial State : $1s^2 2s^1 2p^6 3s^2 3p^6 3d^9 4s^2$ | | | |
|---|---|---|---|
| **Final State** | **$\ell_c$** | **$E_c$ (Ry)** | **$\overline{A_a}$ (Sec⁻¹)** |
| $1s^2 2s^2 2p^5 3s^1 3p^6 3d^9 4s^2$ | 1 | 4.2 81 | 7.478 (11) |
| $1s^2 2s^2 2p^5 3s^2 3p^5 3d^9 4s^2$ | 0 | 8.491 | 2.0984(12) |
| | 2 | | 2.6772 (12) |
| $1s^2 2s^2 2p^5 3s^2 3p^6 3d^8 4s^2$ | 1 | 10.5811 | 7.3189 (11) |
| | 3 | | 2.5865 (12) |
| $1s^2 2s^2 2p^6 3s^2 3p^5 3d^9 4s^1$ | 1 | 9.381 | 2.3046 (10) |
| $1s^2 2s^2 2p^6 3p^6 3d^9 4s^2$ | 0 | 17.545 | 1.06092 (14) |
| $1s^2 2s^2 2p^6 3s^1 3p^5 3d^9 4s^2$ | 1 | 21.206 | 3.4348 (14) |
| $1s^2 2s^2 2p^6 3s^1 3p^6 3d^8 4s^2$ | 2 | 49.171 | 2.3034(14) |
| $1s^2 2s^2 2p^6 3s^1 3p^6 3d^9 4s^1$ | 0 | 49.187 | 1.401 (12) |
| $1s^2 2s^2 2p^6 3s^2 3p^4 3d^9 4s^2$ | 0 | 24.878 | 1.3249 (12) |
| | 2 | | 2.9032 (12) |
| $1s^2 2s^2 2p^6 3s^2 3p^5 3d^9 4s^2$ | 1 | 52.304 | 1.5188 (13) |
| | 3 | | 1.5838 (14) |
| $1s^2 2s^2 2p^6 3s^2 3p^5 3d^9 4s^1$ | 1 | 52.863 | 7.9407 (12) |
| $1s^2 s^2 2p^6 3s^2 3p^6 3d^7 4s^2$ | 0 | 79.732 | 1.363 (12) |
| | 2 | | 8.2855(11) |
| | 4 | | 6.525(13) |
| $1s^2 s^2 2p^6 3s^2 3p^6 3d^8 4s^1$ | 2 | 80.265 | 2.3512 (12) |
| $1s^2 2s^2 2p^6 3s^2 3p^6 3d^9$ | 0 | 80.904 | 5.4336(10) |
| | | | $\Gamma_a$=7.34043 (14) |

**Table 6: Auger transition rates in sec⁻¹ resulting from a 2p-hole in Cu (II) , the energy of the emitted electron (in Ry) and the angular momentum of this electron (c). The number between parentheses is the power of 10.**

| Cu (II) Initial State : $1s^22s^22p^53s^23p^63d^94s^2$ | | | |
|---|---|---|---|
| **Final State** | $\ell_c$ | $E_c$ **(Ry)** | $\overline{A_a}$ **(Sec⁻¹)** |
| $1s^22s^22p^63p^63d^94s^2$ | 1 | 7.0597 | 4.0128 (11) |
| $1s^22s^22p^63s^13p^53d^94s^2$ | 0 | 10.714 | 8.9659 (13) |
| | 2 | | 1.3544(13) |
| $1s^22s^22p^63s^13p^63d^84s^2$ | 1 | 38.167 | 1.5537 (13) |
| | 3 | | 8.816 (12) |
| $1s^22s^22p^63s^13p^63d^94s^1$ | 1 | 38.744 | 2.4429 (10) |
| $1s^22s^22p^63s23p^43d^94s^2$ | 1 | 14.382 | 3.3552 (14) |
| | 3 | | 2.0852 (13) |
| $1s^22s^22p^63s^23p^53d^84s^2$ | 0 | 41.827 | 1.9713 (13) |
| | 2 | | 2.6057 (14) |
| | 4 | | 2.8996 (13) |
| $1s^22s^22p^63s^23p^53d^{10}$ | 0 | 52.054 | 1.2431 (12) |
| | 2 | | 7.281 (12) |
| $1s^22s^22p^63s^23p^53d^74s^2$ | 1 | 69.275 | 6.7376 (12) |
| | 3 | | 1.2092 (14) |
| | 5 | | 4.306 (12) |
| $1s^22s^22p^63s^23p^63d^84s^1$ | 1 | 69.837 | 3.2179 (10) |
| | 3 | | 9.6758 (10) |
| $1s^22s^22p^63s^23p^63d^9$ | 1 | 66.694 | 1.3893(10) |
| | | | $\Gamma_a$=9.86064(14) |

**From Tables 3 and 5 we notice the following points:**

**(1)The dominant contributions to the Auger widths of the 2s – ionized forms of the Cu I and Cu II are obtained from the transitions $\overline{A_a}$ (2s → 3s3s) , $\overline{A_a}$ (2s → 3s3p) , $\overline{A_a}$ (2s → 3s3d) and $\overline{A_a}$ (2s → 3p3d) . The second transition yields the maximum contributions.**

**(2)  The ratio between  $\Gamma_a$'s of the configurations Cu I and Cu II is equal to 1.03603. A similar value was obtained for 1s – ionized atoms, where the difference between the two set of transitions is minimal.**

**The following remarks can be also noticed from Tables 4 and 6:**

**(1)The dominant contributions to the values of the Auger widths, $\Gamma_a$'s, corresponding to the 2p – innershell ionizations of Cu I and Cu II  are delivered by the transition $\bar{A}_a$ (2p  →  3s3p) , $\bar{A}_a$ (2p  →  3p3p) , $\bar{A}_a$ (2p →  3p3d) and $\bar{A}_a$ (2p →  3d3d). The Auger transition rates $\bar{A}_a$ (2p  →  3p3p)  and $\bar{A}_a$ (2p  →  3p3d) are superior.**

**(2)The ratio between the $\Gamma_a$'s, of Cu I and Cu II corresponding to the 2p-transitions is 1.0247, which lies in the same order of magnitude of the rations determined for 1s – hole and 2s – hole transitions.**

## 3.3. M - Shell Auger Transitions

**The emitted Auger electrons in this case can be originated by any M-subshell hole, apart from a 3d one. This is attributed to the fact that no transitions corresponding a 3d-hole are allowed. Thus, only 3s – or 3p – hole decay is possible. The 3s  →  3p3p is energetically forbidden. All possible 3s - Auger transition rates and the corresponding widths of the initial states $1s^2 2s^2 p^6 3s^1 3p^6 3d^{10} 4s^1$  and  $1s^2 2s^2 p^6 3s^1 3p^6 3d^9 4s^2$  of Cu I and Cu II, respectively, are given in Tables 7 and 8**

**Table7: Auger transition rates in sec$^{-1}$ resulting from a 3s-hole in Cu (I), the energy of the emitted electron (in Ry) and the angular momentum of this electron (c). The number between parentheses is the power of 10.**

| Cu (I) Initial State : $1s^2 2s^2 2p^6 3s^1 3p^6 3d^{10} 4s^1$ | | | |
|---|---|---|---|
| **Final State** | $\ell_c$ | **$E_c$ (Ry)** | $\overline{A_a}$ **(Sec$^{-1}$)** |
| $1s^2 2s^2 2p^6 3s^2 3p^5 3d^9 4s^1$ | 1 | 1.981 | 3.1235 (12) |
| | 3 | | 3.832 (12) |
| $1s^2 2s^2 2p^6 3s^2 3p^5 3d^{10}$ | 1 | 2.36 | 5.2230 (11) |
| $1s^2 2s^2 2p^6 3s^2 3p^4 3d^8 4s^1$ | 0 | 8.596 | 5.1768 (12) |
| | 2 | | 7.6765 (13) |
| | 4 | | 3.3715 (14) |
| $1s^2 2s^2 2p^6 3s^2 3p^6 3d^9$ | 2 | 9.759 | 3.22186 (14) |
| | | | $\Gamma_a$=7.448756 (14) |

**Table 8: Auger transition rates in sec$^{-1}$ resulting from a 3s-hole in Cu (II), the energy of the emitted electron (in Ry) and the angular momentum of this electron (c). The number between parentheses is the power of 10.**

| Cu (II) Initial State : $1s^2 2s^2 2p^6 3s^1 3p^6 3d^9 4s^2$ | | | |
|---|---|---|---|
| **Final State** | $\ell_c$ | **$E_c$ (Ry)** | $\overline{A_a}$ **(Sec$^{-1}$)** |
| $1s^2 2s^2 2p^6 3s^2 3p^5 3d^8 4s^2$ | 1 | 1.861 | 2.1376 (12) |
| | 3 | | 2.7878 (12) |
| $1s^2 2s^2 2p^6 3s^2 3p^5 3d^9 4s^1$ | 1 | 2.271 | 3.8861 (11) |
| $1s^2 2s^2 2p^6 3s^2 3p^6 3d^7 4s^2$ | 0 | 8.178 | 4.0304 (12) |
| | 2 | | 6.1038 (13) |
| | 4 | | 3.2957 (14) |
| $1s^2 2s^2 2p^6 3s^2 3p^6 3d^8 4s^1$ | 2 | 8.265 | 2.9462 (14) |
| $1s^2 2s^2 2p^6 3s^2 3p^6 3d^9$ | 0 | 9.2921 | 8.3378 (12) |
| | | | $\Gamma_a$=7.22149 (14) |

**Tables 7 and 8 lead to the following remarks:**

**(1) The two Auger transitions $\bar{A}_a$ (3s → 3d3d) and $\bar{A}_a$ (3s → 4s3d) yield the dominant contributions to the Auger widths of both Cu I and Cu II.**

**(2) The energy of the continuum electron increases as n or ℓ increases for equivalent or nonequivalent effective electrons.**

**(3) The maximum values of $E_c$ are 9.759 Ry for Cu I and 9.2921 Ry for Cu II.**

**(4) In general the continuum energies of the 3s – transitions are smaller than the continuum energies of the previously calculated transitions.**

**(5) The ratio of $\Gamma_a$'s of the two copper configurations is 1.03147 which is nearly equal to the ratios calculated for the 1s – hole, 2s – hole and 2p – hole transitions.**

**The results of 3p – hole decays in Cu I and Cu II are accumulated in Tables 9 and 10. From the Tables we conclude the following points:**

**(1) The number of possible 3p – Auger transitions is reduced to (3p → 3d3d) and (3p → 3d4s). Also the transition (3p →4s4s) is allowed. The first of these three transitions possesses the larges contribution to the widths in both Cu I and Cu II.**

**(2) The Auger transition $\bar{A}_a$ (3p → 3d3d) is the larges transition corresponding to a continuum energy $E_c \approx 65$ eV, which represents the peak of the Auger spectrum.**

**Table 9: Auger transition rates in sec$^{-1}$ resulting from a 3s-hole in Cu (I) , the energy of the emitted electron (in Ry) and the angular momentum of this electron (c). The number between parentheses is the power of 10.**

| Cu (I) Initial State : $1s^2 2s^2 2p^6 3s^2 3p^5 3d^{10} 4s^1$ | | | |
|---|---|---|---|
| **Final State** | $\boldsymbol{\ell_c}$ | $\mathbf{E_c}$ **(Ry)** | $\overline{A_a}$ **(Sec$^{-1}$)** |
| $1s^2 2s^2 2p^6 3s^2 3p^6 3d^8 4s^1$ | 1 | 4.9853 | 3.0550 (14) **Scientific Research and Essays** |
| | 3 | | 4.2263(15) |
| | 5 | | 3.3497(12) |
| $1s^2 2s^2 2p^6 3s^2 3p^6 3d^9$ | 1 | 5.299 | 5.9939 (13) |
| | 3 | | 6.5071(12) |
| | | | $\Gamma_a$=4.601596 (15) |

**Table10: Auger transition rates in sec$^{-1}$ resulting from a 3p-hole in Cu (II) , the energy of the emitted electron (in Ry) and the angular momentum of this electron (c). The number between parentheses is the power of 10.**

| Cu (I) Initial State : $1s^2 2s^2 2p^6 3s^2 3p^5 3d^9 4s^2$ | | | |
|---|---|---|---|
| **Final State** | $\boldsymbol{\ell_c}$ | $\mathbf{E_c}$ **(Ry)** | $\overline{A_a}$ **(Sec$^{-1}$)** |
| $1s^2 2s^2 2p^6 3s^2 3p^6 3d^7 4s^2$ | 1 | 4.4533 | 1.39 (14) |
| | 3 | | 3.8976 (15) |
| | 5 | | 1.2944 (12) |
| $1s^2 2s^2 2p^6 3s^2 3p^6 3d^8 4s^1$ | 1 | 5.155 | 2.707 (14) |
| | 3 | | 1.223 (13) |
| $1s^2 2s^2 2p^6 3s^2 3p^6 3d^9$ | 1 | 5.006 | 6.8177(10) |
| | | | $\Gamma_a$=4.3215093 (15) |